\documentclass{mn2e}
\input{epsf}

\newcounter{subfigure}

\title[The importance of interloper removal]
{The importance of interloper removal in galaxy clusters: saving more objects for the Jeans analysis}
\author[R. Wojtak \& E. L. {\L}okas]{Rados{\l}aw Wojtak and Ewa L. {\L}okas\\
Nicolaus Copernicus Astronomical Center, Bartycka 18, 00-716 Warsaw, Poland}

\begin{document}

\maketitle

\begin{abstract}
We study the effect of contamination by interlopers in kinematic samples of galaxy clusters.
We demonstrate that without the proper removal of interlopers
the inferred parameters of the mass distribution in the cluster are strongly biased towards
higher mass and lower concentration. The interlopers are removed using two procedures previously
shown to work most efficiently on simulated data. One is based on using the virial mass estimator and
calculating the maximum velocity available to cluster members and the other relies on the ratio
of the virial and projected mass estimators. We illustrate the performance of the methods in detail
using the example of A576, a cluster with a strong uniform background contamination, and
compare the case of A576 to 15 other clusters with different degree of contamination. We model the
velocity dispersion and kurtosis profiles obtained for the cleaned data samples of these clusters
solving the Jeans equations to estimate the mass, concentration and anisotropy parameter. We present
the mass-concentration relation for the total sample of 22 clusters.
\end{abstract}


\section{Introduction}

The problem of contamination of kinematic samples of galaxies in clusters by foreground and background
galaxies has been recognized since such data became available. It arises because of the fact that only the
projected positions and velocities of galaxies are measured in redshift surveys. Due to the lack of
knowledge about the motion perpendicular to the line of sight it is difficult to judge a priori
which of the galaxies found close to the cluster in projected space are actually bound to it
and good tracers of the underlying potential. Including unbound galaxies, or
{\em interlopers\/}, in the samples used for
dynamical modelling may lead to significant bias in the estimated parameters of mass distribution
in clusters.

Many procedures have been proposed in the literature to deal with this problem.
All these procedures aim at cleaning the galaxy sample from non-members
before attempting the proper dynamical analysis of the cluster. In the case of single clusters
such a {\em direct\/} approach is the only one possible because of small samples.
Only in the case of kinematic samples combined
from the data for many clusters one can attempt to take the presence of interlopers into account
statistically (van der Marel et al. 2000; Mahdavi \& Geller 2004).

The traditional direct
approach by Yahil \& Vidal (1977) relies on calculation of the velocity dispersion of the galaxy
sample $\sigma$ and iterative removal of outliers with velocities larger than $3 \sigma$.
In rich enough samples one can take into account
the dependence of $\sigma$ on the projected distance from the cluster centre $R$
and perform the rejection
procedure in bins with different $\sigma$ or fit a simple solution of the Jeans equation to the
measured $\sigma (R)$ profile and reject galaxies outside the $3 \sigma (R)$ lines ({\L}okas et al.
2006a). Perea, del Olmo \& Moles (1990) discussed another method relying on iterative removal of
galaxies whose absence in the sample causes the biggest change in the mass estimator.
One can also combine
the information on the position and velocity of a galaxy to infer the maximum velocity available
to cluster members as proposed by den Hartog \& Katgert (1996).

The methods listed above (or variations of them) were recently tested in detail by Wojtak et al.
(2006) using the results of cosmological $N$-body simulations. Working with mock data samples created
from the simulated dark matter haloes they verified what fraction of unbound particles is actually
removed from the sample. They found that the methods of den Hartog \& Katgert (1996) and Perea
et al. (1990) are the most efficient.

{\L}okas et al. (2006a) have recently modelled the kinematics of six nearby regular galaxy clusters
with very few interlopers which could be reliably removed with the simple $3 \sigma (R)$ scheme.
Modelling both the velocity dispersion and kurtosis profiles allowed them to estimate not only
the mass and concentration parameter of the cluster, but also the anisotropy of galactic orbits.
On the opposite end, in terms of the difficulty in estimating the cluster properties, lies A1689.
As discussed by {\L}okas et al. (2006b) the cluster velocity distribution is so complex that no
reliable estimates of its mass and concentration can be obtained from the kinematics.

In this paper we study the kinematics of clusters which could be classified as an intermediate case
between the two mentioned above. We focus on significantly contaminated clusters, for which more
sophisticated methods of interloper removal have to be applied, but still reliable estimates of
the parameters can be obtained. We illustrate the performance of
these methods in detail using the example of a rich nearby ($z=0.04$) galaxy cluster
Abell 576 (section 2). The cluster has a regular image in X-rays which suggests that it is in dynamical
equilibrium. However, contrary to similarly regular clusters like those studied by {\L}okas et al.
(2006a) it is rather strongly contaminated by foreground and background galaxies which are hard to
separate from the main body of the cluster. We show how the inferred parameters of the cluster mass
distribution would be biased if no interloper removal scheme was applied. Then we discuss the
performance of different interloper removal procedures. Finally we model the velocity dispersion
and kurtosis profiles of the cleaned samples to estimate reliably the virial mass, concentration and
anisotropy parameter of the cluster.

In order to place the results for A576 in the right context we study the kinematics of 15 more
nearby clusters showing different degree of contamination (section 3). We demonstrate that after
careful removal of interlopers such clusters can be modelled using Jeans formalism and present
the results of fitting their velocity dispersion and kurtosis profiles in terms of estimated mass,
concentration and anisotropy parameter (section 4). This allows us to extend the mass-concentration
relation obtained from the kinematic data by {\L}okas et al. (2006a) to the total of 22 clusters.

\section{Modelling of A576}

\subsection{Contamination of the data by interlopers}

\begin{figure*}
\begin{center}
    \leavevmode
    \epsfxsize=17cm
    \epsfbox[60 50 700 640]{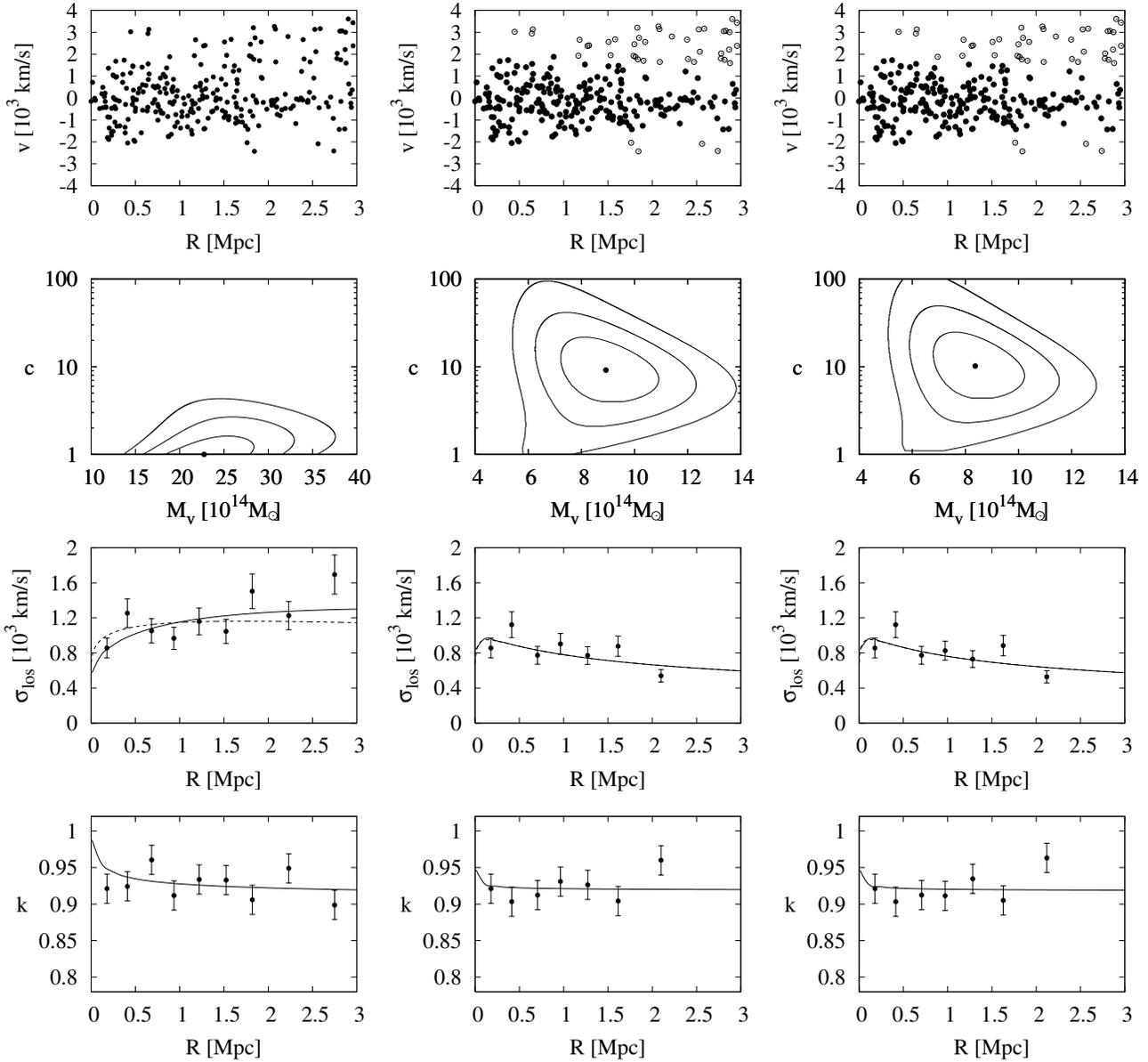}
\end{center}
\caption{First row:
velocity diagrams for the initial sample of galaxies (left) and after removal of
interlopers with the method of $v_{\rm max}$ (middle) and the method using $M_{P}/M_{VT}$ (right).
Rejected galaxies are marked with open circles, those included in the sample with filled circles.
Second row:
results of fitting $\sigma_{\rm los}$ assuming $\beta=0$ in the form of 68.3, 95.4 and 99.73 percent
probability contours in the $M_v-c$ parameter plane. The best-fitting values
of the parameters are indicated with a dot.
Last two rows: velocity dispersion and kurtosis parameter profiles for the corresponding
samples of galaxies. Dashed lines show
the best-fitting $\sigma_{\rm los}$ profiles obtained by fitting $\sigma_{\rm los}$ data
alone assuming $\beta=0$, solid lines
show profiles from the joint fitting of $\sigma_{\rm los}$ and $\kappa_{\rm los}$ for arbitrary orbits.}
\label{comparison}
\end{figure*}

The positions and redshifts of galaxies in the vicinity of A576 were obtained
from the NASA/IPAC Extragalactic Database (NED). We have searched the database for
galaxies within projected distance of 3 Mpc from the cluster centre and
with $c z$ velocities differing from the cluster mean $z=0.0389$
by less than $\pm 4000$ km s$^{-1}$. The redshifts of galaxies thus selected come mainly from
Mohr et al. (1996) and Rines et al. (2000). The cluster possesses two central galaxies. We choose one
of these galaxies with redshift closer to the cluster mean as the centre of the cluster, with
respect to which all distances are measured.

The presence of this central galaxy pair may suggest
that the cluster underwent a merger. This hypothesis is also supported by X-ray data:
although the large scale X-ray surface brightness
distribution measured by Einstein (Mohr et al. 1996) and ASCA satellites (Horner et al. 2000)
is rather regular, some departures from
equilibrium in the cluster core have been detected with Chandra (Kempner \& David 2004;
Dupke \& Bregman 2005).
However, since the X-ray map is not strongly perturbed and the centre of the gas distribution
coincides with the position of the two central galaxies
the cluster probably had time to reach equilibrium after the merging event.

The galaxy velocities have been transformed to the reference frame of the cluster and in order to
calculate the distances we have transformed the cluster velocities to the
reference frame of the cosmic microwave background and assumed the $\Lambda$CDM cosmological model
with parameters $\Omega_M=0.3$, $\Omega_{\Lambda}=0.7$ and $h=0.7$. The line-of-sight
velocities of 272 galaxies
as a function of their projected distance from the cluster centre are plotted in the upper left
panel of Fig.~\ref{comparison}. We will refer to this kind of plot as a velocity diagram.
We can see that although the main body of the cluster is well visible in the diagram, there is
also a strong contamination by foreground and background galaxies which are unlikely members of
the cluster.

The third panel in the left column of Fig.~\ref{comparison} shows the velocity dispersion profile
$\sigma_{\rm los}$ calculated from the sample with the standard unbiased dispersion estimator
and $n=30$ galaxies per bin. The data points were assigned sampling errors of size
$\sigma_{\rm los}/\sqrt{2(n-1)}$.
We can see that the velocity dispersion profile increases strongly with the projected distance $R$
contrary to what is expected for a relaxed galaxy cluster. For such objects, real or simulated, the
density profile is close to NFW (Navarro, Frenk \& White 1997) while orbits do not depart
significantly from isotropic and the line-of-sight velocity dispersion profile $\sigma_{\rm los} (R)$
should be a decreasing function of $R$ ({\L}okas \& Mamon 2001). We can try to reproduce this
$\sigma_{\rm los} (R)$ with the solution of the Jeans equation (Binney \& Mamon 1982)
\begin{equation}           \label{sigma_los}
	\sigma_{\rm los}^2 (R)=\frac{2}{\Sigma_{M} (R)}\int_{R}^{\infty}\frac{\rho(r)
	\sigma^{2}_{r}r}{\sqrt{r^{2}-R^{2}}}\Big(1-
	\beta\frac{R^{2}}{r^{2}}\Big)\textrm{d}r,
\end{equation}
where $\rho(r)$ is the NFW
density profile and $\Sigma_{M}(R)$ its two-dimensional projection, assuming
the anisotropy parameter $\beta=1-\sigma^{2}_{\theta}(r)/\sigma^{2}_{r}(r)=0$ and estimating
the virial mass $M_v$ and concentration parameter $c$ (we assume $c>1$). The results of the
fitting procedure are shown in the second panel in the left column of Fig.~\ref{comparison}
in terms of the
68.3, 95 and 99.73 percent probability contours in the $M_v-c$ parameter plane. We see that
the preferred concentration is the lowest possible ($c=1$) and the virial mass is as high as
$M_v = 2.3 \times 10^{15}$ M$_{\sun}$. The solution is plotted as a dashed line in the third
panel of the left column of the Figure.

The final modelling of galaxy cluster kinematics should take into account possible departures
from isotropy of galactic orbits. Since the velocity dispersion data alone cannot break the degeneracy
between the anisotropy and the shape (concentration) of the density profile it is useful to
consider a higher order velocity moment, the line-of-sight kurtosis
$\kappa_{\rm los} (R) = \overline{v_{\rm los}^4} (R)/\sigma_{\rm los}^4 (R)$
({\L}okas \& Mamon 2003; {\L}okas et al. 2006a). The kurtosis profile expressed in terms of
$k = \left[ \log \left( 3 K/2.68 \right) \right]^{1/10}$ (where $K$ is the standard
kurtosis estimator) for the total sample of galaxies in A576 is shown in the lower left panel
of Fig.~\ref{comparison}. Note that contrary to the common belief that the higher moments
of velocity should be more affected by interlopers (which is indeed often the case)
the kurtosis profile is quite
flat and similar to profiles for regular clusters with very low contamination
(see Fig. 5 in {\L}okas et al. 2006a).

\begin{table}
\begin{center}
\caption{Fitted parameters of A576 with $1\sigma$ error bars.
The first three rows show the results obtained from
fitting only $\sigma_{\rm los}$ for the samples obtained with simple velocity cut-off,
the method of $v_{\rm max}$ and
the method based on $M_{P}/M_{VT}$ assuming isotropic orbits. The last three rows list the
parameters obtained from joint fitting of $\sigma_{\rm los}$ and $\kappa_{\rm los}$ for arbitrary
orbits.}
\begin{tabular}{ccccc}
method & $\beta$ & $M_v [10^{14} M_{\sun}]$               & $c$  & $\chi^2/N$ \\
\hline
cut-off        & 0                      & $23^{+5}_{-5}$      & $1.0^{+0.8}_{-0.0}$  & 9.4/7   \\  \\
$v_{\rm max}$  & 0                      & $8.9^{+1.9}_{-1.5}$ & $9.2^{+12.0}_{-5.0}$  & 9.4/5   \\  \\
$M_{P}/M_{VT}$ & 0                      & $8.4^{+1.9}_{-1.6}$ & $10.2^{+15.0}_{-4.7}$ & 9.2/5   \\
\hline
cut-off        & $-0.75^{+0.85}_{-1.55}$& $40^{+43}_{-20}$    & $1.0^{+2.0}_{-0.0}$   & 10.6/15   \\  \\
$v_{\rm max}$  & $0.11^{+0.64}_{-1.51}$ & $9.0^{+2.3}_{-2.2}$ & $8.6^{+18.0}_{-5.6}$  & 15.6/11 \\  \\
$M_{P}/M_{VT}$ & $0.07^{+0.73}_{-1.52}$ & $8.4^{+2.3}_{-2.0}$ & $10.0^{+22.0}_{-7.0}$ & 16.4/11 \\
\hline
\label{fittedpar}
\end{tabular}
\end{center}
\end{table}

Performing a joint fitting of velocity dispersion and kurtosis profiles to the solutions of the
Jeans equations ({\L}okas \& Mamon 2003; {\L}okas et al. 2006a) assuming $\beta=$ const
we arrive at an even higher
mass estimate of $M_v=4.0 \times 10^{15}$ M$_{\sun}$ (again with $c = 1$) and anisotropy parameter
$\beta=-0.75$. The parameters are summarized together with $1\sigma$ error bars in Table~\ref{fittedpar}
in the lines labelled as `cut-off' referring to the velocity cut-off by which the galaxy sample
was selected.
The best-fitting solutions are plotted in the lower left panels of Fig.~\ref{comparison} as solid lines.
The reason for the higher mass estimate in the case of fitting both velocity moments
is the following: since kurtosis is not strongly overestimated
it allows tangential orbits (see Fig. 10 in Sanchis, {\L}okas \& Mamon 2004) which also fit the dispersion
profile better, but for a higher mass.

\subsection{Removal of interlopers}

The method of interloper removal found to work best on simulated data (see Table~1 of
Wojtak et al. 2006) was proposed by den Hartog
\& Katgert (1996). This approach was successful in removing on average 73 percent of unbound
particles from simulated velocity diagrams obtained with the same initial cut-off in line-of-sight
velocity, as applied here to A576: $\pm 4000$ km s$^{-1}$ with respect to the cluster mean.
The unbound particles remaining in the sample afterwards were those falling within the cluster
main body in the projected phase space. Those remaining interlopers do not bias the velocity
dispersion significantly.

The method relies on calculating the maximum velocity available to cluster members. The velocity
is estimated assuming that a galaxy is either on a circular orbit with velocity
$v_{\rm cir} = \sqrt{GM(r)/r}$ or infalling into the cluster with velocity $v_{\rm inf} =
\sqrt{2}v_{\rm cir}$ so that
\begin{equation}	\label{vmax}
	v_{\rm max}={\rm max}_{R}\{v_{\rm inf} \cos\theta, v_{\rm cir} \sin\theta \} ,
\end{equation}
where $\theta$ is the angle between position vector of the galaxy with respect
to the cluster centre and the line of sight.

In order to calculate the maximum velocity we need an estimate of the mass profile.
It turns out (see Wojtak et al. 2006) that the method works best if this is done with
the mass estimator $M_{VT}$ derived from the virial
theorem (Heisler, Tremaine \& Bahcall 1985)
\begin{equation}	\label{M_VT}
	M_{VT}(r=R_{\rm max})=\frac{3\pi N}{2G}\frac{\Sigma_{i}
	(v_{i}-\bar{v})^{2}}{\Sigma_{i<j}1/R_{i,j}} ,
\end{equation}
where $N$ is a number of galaxies with $R<R_{\rm max}$, $v_{i}$ is the velocity of the $i$-th
galaxy and $R_{i,j}$ is a projected distance between $i$-th and $j$-th galaxy.
The mass profile can be then obtained from $M(r)\approx
M_{VT}(R_{i}<r<R_{i+1})$, where $R_{i}$ is the sequence of projected radii of
galaxies in the increasing order up to $R_{\rm max} =3$ Mpc in our case. From this mass
profile we calculate the maximum velocity profile (\ref{vmax})
and remove galaxies with velocities exceeding $v_{\rm max}$. The procedure is
repeated until no more interlopers are removed.

\begin{figure}
\begin{center}
    \leavevmode
    \epsfxsize=8cm
    \epsfbox[55 55 395 295]{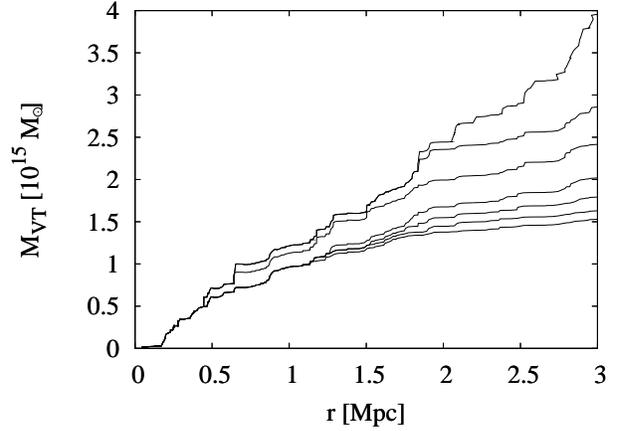}
\end{center}
\caption{Mass profiles estimated from eq. (\ref{M_VT}) in the iterative procedure of interloper
removal based on rejection of galaxies with velocities larger than $v_{\rm max}$ given by
eq. (\ref{vmax}). The highest line is for the first iteration, the lowest for the last.}
\label{hartog}
\end{figure}

The performance of the method in the case of A576 is illustrated in Fig.~\ref{hartog}
showing the mass profiles in subsequent iterations with the highest line corresponding to
the first iteration and the lowest to the last one. While in the first iteration the mass
within $R_{\rm max} =3$ Mpc is as high as $4 \times 10^{15}$ M$_{\sun}$, in the last iteration
it decreases down to $1.5 \times 10^{15}$ M$_{\sun}$. The resulting final sample of galaxies
is shown in the upper middle panel of Fig.~\ref{comparison} as filled circles, while the
rejected galaxies were marked as open circles. The corresponding velocity dispersion profile
is plotted in the third middle panel of Fig.~\ref{comparison} (we restrict the analysis to seven
inner data points so that all contributing galaxies are within the estimated virial radius).
It is now a decreasing function of
the projected radius $R$, as expected for samples free of contamination. The best-fitting NFW
parameters obtained from fitting $\sigma_{\rm los}$
are now $M_v = 8.9 \times 10^{14}$ M$_{\sun}$ and $c=9.2$ with the confidence
regions shown in the second middle panel of Fig.~\ref{comparison}.

Allowing for arbitrary constant anisotropy and adding kurtosis to the analysis (lower middle
panel of Fig.~\ref{comparison}) we arrive at very similar numbers, in particular the estimated
anisotropy parameter is very close to zero (see Table~\ref{fittedpar}). The best-fitting profiles
of velocity moments in this case are shown in the middle column of Fig.~\ref{comparison} with
solid lines. Note that the
$1\sigma$ error bars estimated from joint fitting of both velocity moments, but keeping
$\beta$ as a free parameter, are larger than those found from fitting $\sigma_{\rm los}$ alone
assuming $\beta=0$.

Another method of interloper removal that we find useful in the case of A576 is the one based
on the use of mass estimators (see Table~1 of Wojtak et al. 2006). In this
method, originally proposed by Perea et al. (1990), in addition to the virial mass estimator
(\ref{M_VT}) we consider the projected mass estimator which for $\beta=0$ reads
\begin{equation}	\label{M_P}
	M_{P}=\frac{32}{\pi GN}\Sigma_{i}(v_{i}
	-\bar{v})^{2}R_{i}
\end{equation}
where again $v_{i}$ and $R_{i}$ are velocity and projected radius of the $i$-th
galaxy and $N$ is the number of galaxies in the sample. As discussed by Wojtak et al. (2006)
the presence of interlopers causes strong overestimation of both $M_{VT}$ and $M_{P}$, but more
so in the case of $M_{P}$, one can therefore construct a method of interloper removal based on
the ratio $M_{P}/M_{VT}$. Such a method is able to remove on average 65 percent of unbound
galaxies from velocity diagrams.

The method uses the jackknife statistics in order to eliminate the interlopers. The data are arranged
in $\{R_{i},v_{i}\}$ sequence, where $i$ goes from 1 to $n$ and we calculate $n$ values of both mass
estimators corresponding to $n$ subsequences with one data point excluded.
We identify as an interloper the galaxy for which the corresponding
subsequence is the source of the most discrepant value of one of the estimators.
In the next step, the same procedure is applied to a new data set with $n-1$ particles.
The disadvantage of this procedure is that it does not converge and we have to define some
criterion for stopping the algorithm. Since in the case of samples cleaned from interlopers
we should have $M_{P}/M_{VT}\rightarrow 1$ and $\Delta M/M\rightarrow 0$ these would be the
most obvious conditions. Unfortunately, in the case of simulated clusters they are never exactly
fulfilled and more conservative relations have to be adopted.

\begin{figure}
\begin{center}
    \leavevmode
    \epsfxsize=8cm
    \epsfbox[55 55 395 545]{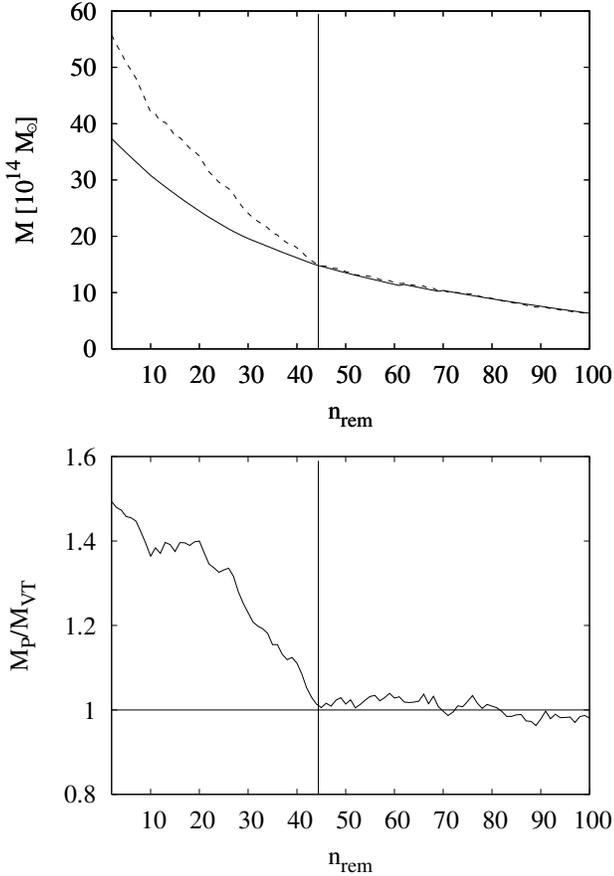}
\end{center}
\caption{Upper panel: the virial mass estimator $M_{VT}$ (solid line) and the projected mass
estimator $M_{P}$ (dashed line) as a function of the number of removed galaxies $n_{\rm rem}$.
Lower panel: the ratio of the two mass estimators $M_{P}/M_{VT}$ as a function of $n_{\rm rem}$.
Vertical lines in both panels mark $n_{\rm rem}=44$ when the procedure should be stopped.}
\label{jackknife}
\end{figure}

However, applying the algorithm to the kinematic sample of A576 we discovered that the condition
$M_{P}/M_{VT}\rightarrow 1$ is actually fulfilled after some interlopers are removed.
The performance of the procedure is illustrated in Fig.~\ref{jackknife}, where we plot the values
of the two estimators (upper panel) and their ratio (lower panel) as a function of the number
of removed galaxies $n_{\rm rem}$. We find that at $n_{\rm rem} = 44$ the ratio $M_{P}/M_{VT}$
drops to unity
and remains at this level even if more galaxies are rejected. This moment is a clear signature that
the sample has been cleaned of interlopers.

The final sample of galaxies from this method after removal of $n_{\rm rem} = 44$ galaxies
is shown in the upper right panel of Fig.~\ref{comparison}. The galaxies accepted by the
procedure are plotted with filled circles, while those removed with open circles. The corresponding
velocity dispersion profile and the constraints on the parameters from the solution of the
Jeans equation with $\beta=0$ are also
shown in the right column of the Figure. The best-fitting parameters are now
$M_v = 8.4 \times 10^{14}$ M$_{\sun}$ and $c=10.2$. Adding kurtosis to the analysis (lower right
panel of Fig.~\ref{comparison}) we again arrive at very similar numbers
(see Table~\ref{fittedpar}). The best-fitting profiles
of velocity moments in this case are shown in the right column of Fig.~\ref{comparison} with
solid lines.

The parameters estimated from fitting
velocity dispersion and both moments for the samples
obtained by the two methods are summarized in Table~\ref{fittedpar}.
The present final sample differs from the one obtained with the previous method only by one
galaxy which is removed here and was retained in the previous case. This single galaxy is however
responsible for the difference between the velocity dispersion and kurtosis profiles in the two cases
(we keep 30 galaxies per bin so the binning of the data is different). The fitted parameters of the
NFW mass distribution however remain quite similar.

Interestingly, no other method studied by Wojtak et al. (2006) and shown to work reasonably well for
simulated clusters, is applicable to A576. In particular, the commonly used $3 \sigma$ method
originally proposed by Yahil \& Vidal (1977) does not remove any galaxy from the sample, whether
applied to a single bin or in radial bins of different size. Similarly, the method based on fitting
the isotropic solution of the Jeans equation and rejecting galaxies outside the $3 \sigma_{\rm los}
(R)$ lines successfully applied to less contaminated clusters by {\L}okas et al. (2006a) does not
work here (it removes only one galaxy).
The reason is that in the case of A576 we do not have single outliers but rather a
uniform background of interlopers. This makes the dispersion profile for the contaminated
sample increase
so strongly (see the left column of Fig.~\ref{comparison}) that all galaxies fall within the
fitted $3 \sigma_{\rm los} (R)$ lines and the procedure does not start.

\subsection{Comparison with earlier work}

We studied the performance of a few methods of interloper removal, two of which work especially well:
one based on estimating the maximum velocity available for cluster members and the other on
the use of the ratio of two mass estimators. The methods work equally well producing reliable
final galaxy samples that lead to similar conclusions about the mass distribution in the cluster.
If we were to recommend one of these methods, we would choose the first since on average it
should remove more unbound galaxies and is always convergent.

Other methods of selecting the cluster members have been applied to A576 before. For example,
Mohr et al. (1996) used galaxies without emission lines as likely virialized members. However,
the velocity dispersion profile for this sample (see the upper panel of their fig. 8) is still
rather flat at larger distances from the cluster's centre leading to an overestimated mass of
$1.5 \times 10^{15}$ M$_{\sun}$ (while the dispersion of the 20 brightest members of this sample
gives a value twice smaller). This shows that selecting early-type galaxies is not sufficient
to remove interlopers since some of these galaxies can still come from the cluster's neighbourhood.
Rines et al. (2000) used the caustic method to study the infall patterns around A576 using a similar
initial kinematic sample as the one used here. Their selection of members yields a similar sample
as we obtain using our two methods of interloper removal.

Since we believe
the $v_{\rm max}$ method to be more reliable we adopt its results as final:
$M_v=9.0^{+2.3}_{-2.2} \times 10^{14}$ M$_{\sun}$ and $c = 8.6^{+18.0}_{-5.6}$. Our mass estimate
is in excellent agreement with the result of Rines et al. (2000). Within our virial radius
of 2.5 Mpc they find (see their fig.~7) the
projected mass $M_p = (10.1 \pm 1.1) \times 10^{14}$ M$_{\sun}$ which for our best-fitting concentration
corresponds to $M_v = (8.5 \pm 1.0) \times 10^{14}$ M$_{\sun}$. We also confirm their finding that
the mass determined directly from traditional mass estimator $M_{VT}$ strongly overestimates
the true value. As shown in Fig.~\ref{hartog} the last iteration (lowest line) gives at 2.5 Mpc
$M_{VT} = 14.5 \times 10^{14}$ M$_{\sun}$ which is 1.6 times higher than our best estimate from velocity
moments. The same is true for the projected mass estimator $M_P$ which at the moment of stopping
the interloper removal procedure is equal to $M_{VT}$ (see Fig.~\ref{jackknife}). In agreement
with Biviano et al. (2006) we therefore find that these mass estimators overestimate the cluster
mass even if the interlopers are properly removed.

Comparison with mass estimates derived from the condition of hydrostatic
equilibrium of the X-ray emitting gas is more complicated because of a large number of
parameters involved in such analyses. Using the result for
$M_X(<r)$ from Mohr et al. (1996, their equation 5.3) with $T=4$ keV (Kempner \& David 2004)
and $r=2.5$ Mpc we get $M_X(<2.5) \approx 7 \times 10^{14}$ M$_{\sun}$, which is consistent with our mass.
(Restricting the analysis to $r=0.7$ Mpc to avoid the extrapolation of X-ray data we get
$M_X(<0.7) \approx 2 \times 10^{14}$ M$_{\sun}$ while our best estimate of mass within this radius is
$M = 3.4 \times 10^{14}$ M$_{\sun}$.)
However, more detailed modelling of the X-ray data by Rines et al. (2000) showed that the mass
profile inferred from the properties of X-ray gas is systematically lower than
the one derived from galaxy kinematics, as is the case for many clusters. Recently, Benatov et al.
(2006) proposed that this discrepancy can be reduced if the galaxy orbits are radially
anisotropic. Indeed, fixing $\beta$ at the upper limit of the $1\sigma$ confidence regions listed
in the last two lines of Table~\ref{fittedpar} we find that the best-fitting $M_v$ and $c$ are
pushed towards the lower limits of the ranges given in the Table. We therefore confirm that
significantly lower mass estimates are preferred for radial orbits.

\section{Other clusters}

\renewcommand{\thefigure}{\arabic{figure}\alph{subfigure}}
\setcounter{subfigure}{1}

\setcounter{subfigure}{1}
\begin{figure*}
\begin{center}
    \leavevmode
    \epsfxsize=17cm
    \epsfbox[50 50 900 800]{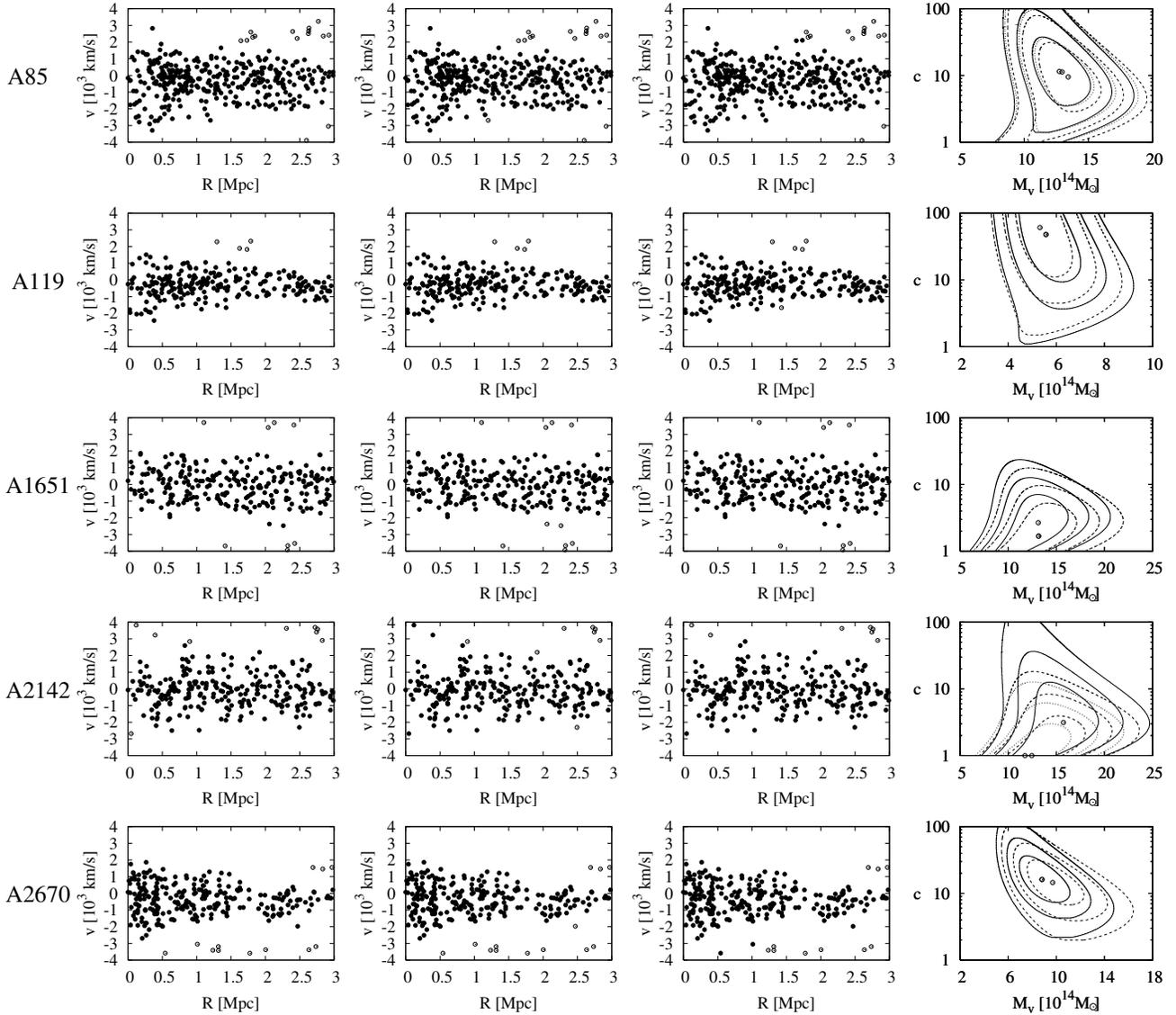}
\end{center}
\caption{Results of interloper removal for the first group of clusters in Table~\ref{clusters_int}.
Three first columns from the left to the right show results of the three methods of interloper removal:
$3\sigma(R)$, $v_{\rm max}$ and $M_{P}/M_{VT}$, respectively.
Filled circles indicate cluster members and open circles the galaxies identified as interlopers.
The right column shows results of fitting $\sigma_{\rm los}$ to the solution of the Jeans equation
(\ref{sigma_los}) with $\beta=0$ in terms of 68.3, 95.4 and 99.73 percent probability contours
in the $M_v-c$ parameter plane. Dotted, solid and dashed lines correspond to samples of galaxies
obtained with $3\sigma(R)$, $v_{\rm max}$ and $M_{P}/M_{VT}$ procedures of interloper removal.
In the cases where only two types of contours are shown in the right column, two methods gave
identical results.}
\label{clusters1}
\end{figure*}

We have shown that for strongly contaminated kinematic samples, as
in the case of A576, the proper removal of interlopers is essential to avoid bias in the
estimated parameters of the mass distribution. One may ask to what extent A576 is special and
whether we should worry about such cases or perhaps just discard them from our cluster samples.
As discussed in {\L}okas et al. (2006a) there are now few tens
nearby ($z < 0.1$) clusters with significant number of galaxy redshifts
available from NED. About half of these clusters have irregular
X-ray images so are far from equilibrium and therefore are not tractable by Jeans
modelling. In addition, some of the clusters, is spite of
regular X-ray images, have close neighbours or strong
substructure which makes the analysis difficult. A576 is a
special case among these clusters in a sense
that it has a strong contamination of interlopers but
otherwise is quite regular.

\addtocounter{figure}{-1}
\addtocounter{subfigure}{1}
\begin{figure*}
\begin{center}
    \leavevmode
    \epsfxsize=17cm
    \epsfbox[50 50 900 800]{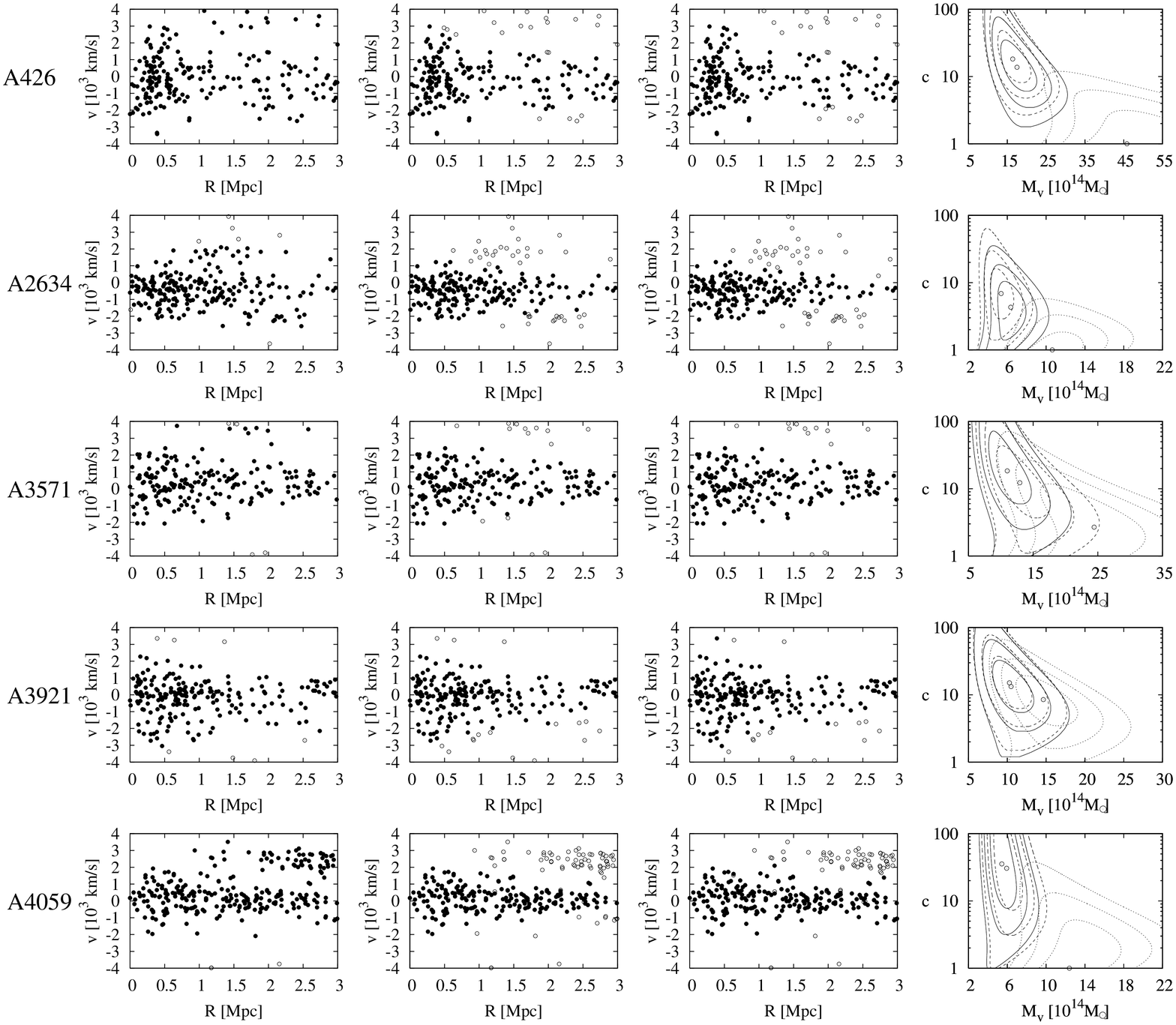}
\end{center}
\caption{The same as Fig.~\ref{clusters1} for the second group of clusters in Table~\ref{clusters_int}.}
\label{clusters2}
\end{figure*}

A576 is not the only cluster which can be retained for dynamical analysis after careful
interloper removal. To illustrate further the importance of interloper removal and
show broader impact of the schemes studied in Wojtak et al. (2006)
we studied 15 more nearby ($z<0.1$) and possibly relaxed (in terms of regularity of their
X-ray maps and
velocity diagrams) clusters with at least 120 galaxies available within projected radius of 2 Mpc.
We apply to all of them different procedures of interloper removal and compare corresponding
samples of cluster galaxies in terms of numbers of rejected galaxies and results of fitting the isotropic
solution of the Jeans equation. For clarity
we focus on three most representative methods described in detail in the previous
sections: $3\sigma(R)$ rejection, $v_{\rm max}$ and $M_{P}/M_{VT}$ (see Wojtak
et al. 2006), since other approaches are very similar. Like for A576, we use
NED to select positions and redshifts of galaxies. Projected distances and the line-of-sight
velocities of galaxies were obtained following the same steps as described in section 2 for A576.
Table~\ref{clusters_int} lists the redshifts, the total numbers of galaxies used
and the results of interloper removal for the 15 selected clusters. The Table also gives the
references to the most important sources of redshift data.

\begin{table}
\begin{center}
\caption{Properties of the cluster sample composed of 15 clusters.
The columns list the cluster name, redshift, total number
of galaxies $N_{\rm tot}$ in the initial sample
and numbers of interlopers identified with three methods of interloper removal.
The boldfaced numbers indicate interlopers excluded for the final modelling.
The last column gives the main sources of the redshift data: 1 - Durret et al. (1998),
2 - Katgert et al. (1998), 3 - Smith et al. (2004), 4 - 2dF Galaxy Redshift Survey
(Colless et al. 2001), 5 - SDSS (Abazajian et al. 2005), 6 - SDSS (Abazajian et al. 2004),
7 - Chincarini \& Rood (1971), 8 - Lucey et al. (1997), 9 - Scodeggio et al. (1995),
10 - Ferrari et al. (1998), 11 - Pimbblet et al. (2006), 12 - Hill \& Oegerle (1993),
13 - Barmby \& Huchra (1998), 14 - Wegner et al. (1999).
}
\begin{tabular}{lcccccl}
cluster & $z$ & $N_{\rm tot}$ & $3\sigma(R)$ & $v_{\rm max}$ & $M_{P}/M_{VT}$ & Ref.\\
\hline
A85    & 0.0551 &  358 & 15   &    \bf{16}   &  13       &  1    \\
A119   & 0.0442 &  239 & 4    &    \bf{4}    &  5        &  2,3  \\
A1651  & 0.0849 &  250 & 8    &    \bf{10}   &  8        &  4    \\
A2142  & 0.0909 &  255 & 9    &    \bf{8}    &  7        &  5    \\
A2670  & 0.0762 &  263 & 12   &    \bf{13}   &  10       &  6    \\
\hline
A426   & 0.0179 &  203 & 0    &    \bf{21}   &  20       &  7    \\
A2634  & 0.0314 &  230 & 7    &    \bf{39}   &  44       &  8,9   \\
A3571  & 0.0391 &  197 & 4    &    \bf{14}   &  12       &  3     \\
A3921  & 0.0928 &  211 & 7    &    \bf{16}   &  14       &  10,11  \\
A4059  & 0.0475 &  279 & 3    &    \bf{70}   &  72       &  3,4   \\
\hline
A401   & 0.0737 &  156 & 0    &    \bf{4}    &  10       &  12    \\
A2147  & 0.0350 &  384 & 3    &    52        &  \bf{109} &  13    \\
A2593  & 0.0413 &  259 & 13   &    \bf{27}   &  15       &  6,14  \\
A3667  & 0.0556 &  197 & 0    &    \bf{5}    &  1        &  3     \\
A4038  & 0.0300 &  315 & 4    &    26        &  \bf{72}  &  4     \\
\hline
\label{clusters_int}
\end{tabular}
\end{center}
\end{table}

We have grouped the clusters in Table~\ref{clusters_int} according to their behaviour
in terms of interloper removal. The results of interloper
removal for the first five clusters in Table~\ref{clusters_int}
are illustrated in Fig.~\ref{clusters1}.
In the first three columns of Fig.~\ref{clusters1} we put velocity diagrams
with filled and open circles indicating cluster galaxies and identified interlopers. Each row
corresponds to a different cluster as marked in the Figure.
The columns from the left to the right correspond respectively to the results of $3\sigma(R)$,
$v_{\rm max}$ and $M_{P}/M_{VT}$ procedures. The last column shows results of fitting isotropic
solution of the Jeans equation, (\ref{sigma_los}) with $\beta=0$, to the velocity dispersion profiles
in the form of 68.3, 95.4 and 99.73 percent probability contours in $M_{v}-c$ plane.
Dotted, solid and dashed lines correspond to samples of cluster galaxies obtained with
$3\sigma(R)$, $v_{\rm max}$ and $M_{P}/M_{VT}$ methods. Similar plots for the two other groups of
clusters listed in Table~\ref{clusters_int} are shown in Fig.~\ref{clusters2} and \ref{clusters3}.

\addtocounter{figure}{-1}
\addtocounter{subfigure}{1}
\begin{figure*}
\begin{center}
    \leavevmode
    \epsfxsize=17cm
    \epsfbox[50 50 900 800]{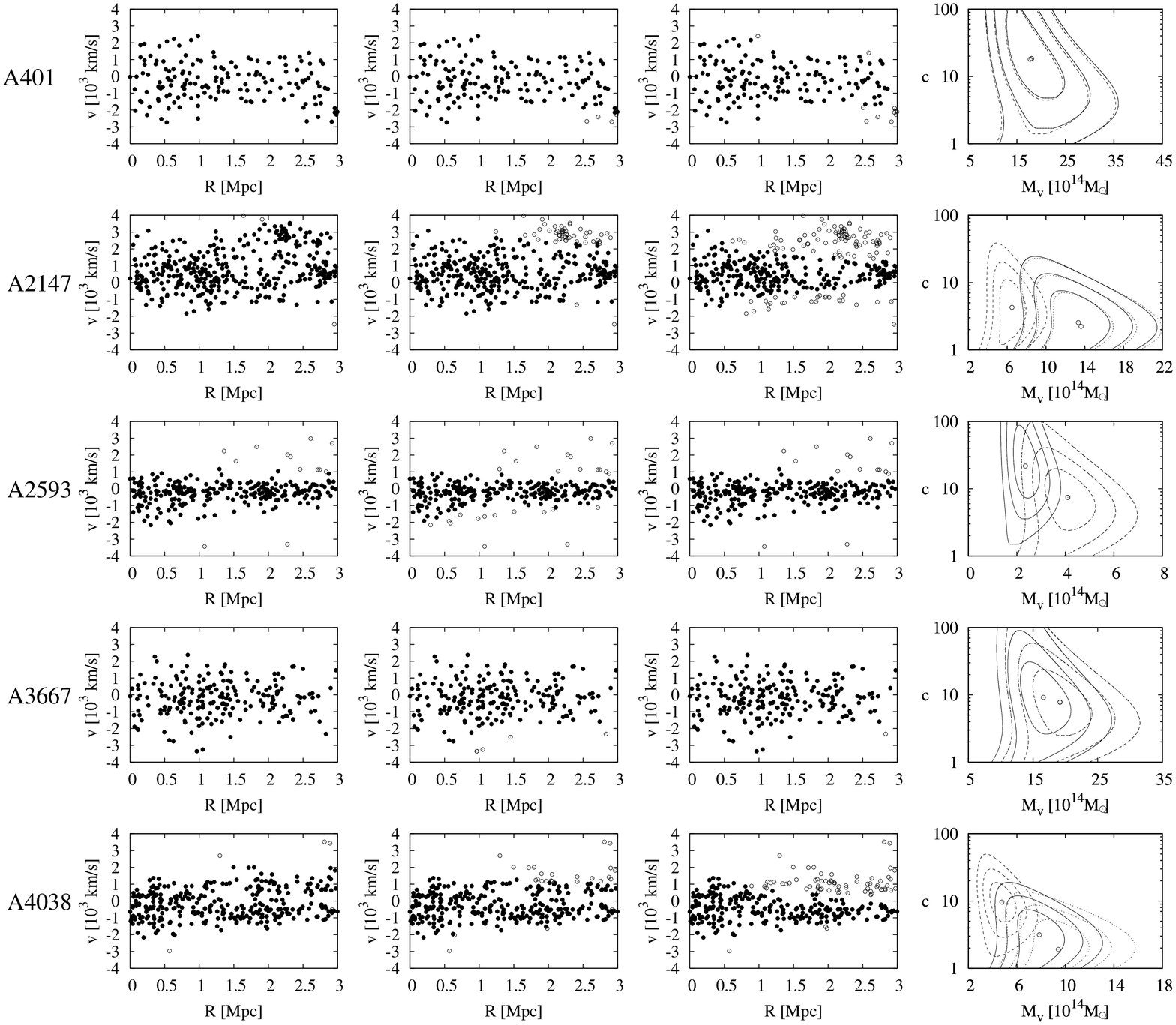}
\end{center}
\caption{The same as Fig.~\ref{clusters1} for the last group of clusters in Table~\ref{clusters_int}.}
\label{clusters3}
\end{figure*}

\renewcommand{\thefigure}{\arabic{figure}}

Since each cluster is different it is difficult to divide them into distinct classes from the point
of view of their behaviour under interloper removal schemes. However,
comparing the numbers of interlopers determined with the three methods (Table~\ref{clusters_int})
we find that for 5 out of 15 clusters (the first group in Table~\ref{clusters_int})
all results of interloper removal are nearly the same.  A good example of this class of clusters is
the A85 shown in the first line of plots in Fig.~\ref{clusters1}.
We notice that in this case only the $M_{P}/M_{VT}$ procedure
seems to give slightly less consistent output. This happens due to the fact that the number of
interlopers here is relatively small and it is hard to find precisely a characteristic knee-like
point in the $M_{P}/M_{VT}$ profile (see Fig.~\ref{jackknife} for A576) at which the algorithm
of interloper removal should be stopped (Wojtak et al. 2006). However, the differences between
the final samples of cluster galaxies are not significant and, consequently, probability contours in
the $M_{v}-c$ plane determined from the data sets overlap each other.

Five clusters of the second group in Table~\ref{clusters_int} illustrate the case when
only $v_{\rm max}$ and $M_{P}/M_{VT}$ schemes
determine samples of cluster galaxies consistent with each other, whereas the $3\sigma(R)$ approach
fails giving rather discrepant results. This situation manifests itself clearly in terms of the
fitting results. Probability contours associated with $v_{\rm max}$ and $M_{P}/M_{VT}$ schemes
coincide almost exactly, whereas the ones corresponding to the $3\sigma(R)$ procedure are shifted towards
higher mass and lower concentration. Note that from the point of view of interloper removal
A576 is very similar to the clusters of this group.

Finally, for the last 5 clusters listed in Table~\ref{clusters_int} all methods of interloper removal
identify different numbers of interlopers. This implies that probability contours associated
with the selected samples of cluster galaxies typically do not coincide as well as in the
case of clusters of
the first section of Table~\ref{clusters_int}. Among these clusters A2147 and A4038
have distinct neighbouring substructures which overlap partially with the main
body of the cluster. This causes the failure of both $3\sigma(R)$ and $v_{\rm max}$ procedures.
Nevertheless, the results obtained with the $M_{P}/M_{VT}$ scheme are likely reliable for these two
clusters since
the moment of convergence of the algorithm is clearly seen in the $M_{P}/M_{VT}$ profile.

\section{Discussion}

\begin{figure*}
\begin{center}
    \leavevmode
    \epsfxsize=17cm
    \epsfbox[60 50 820 430]{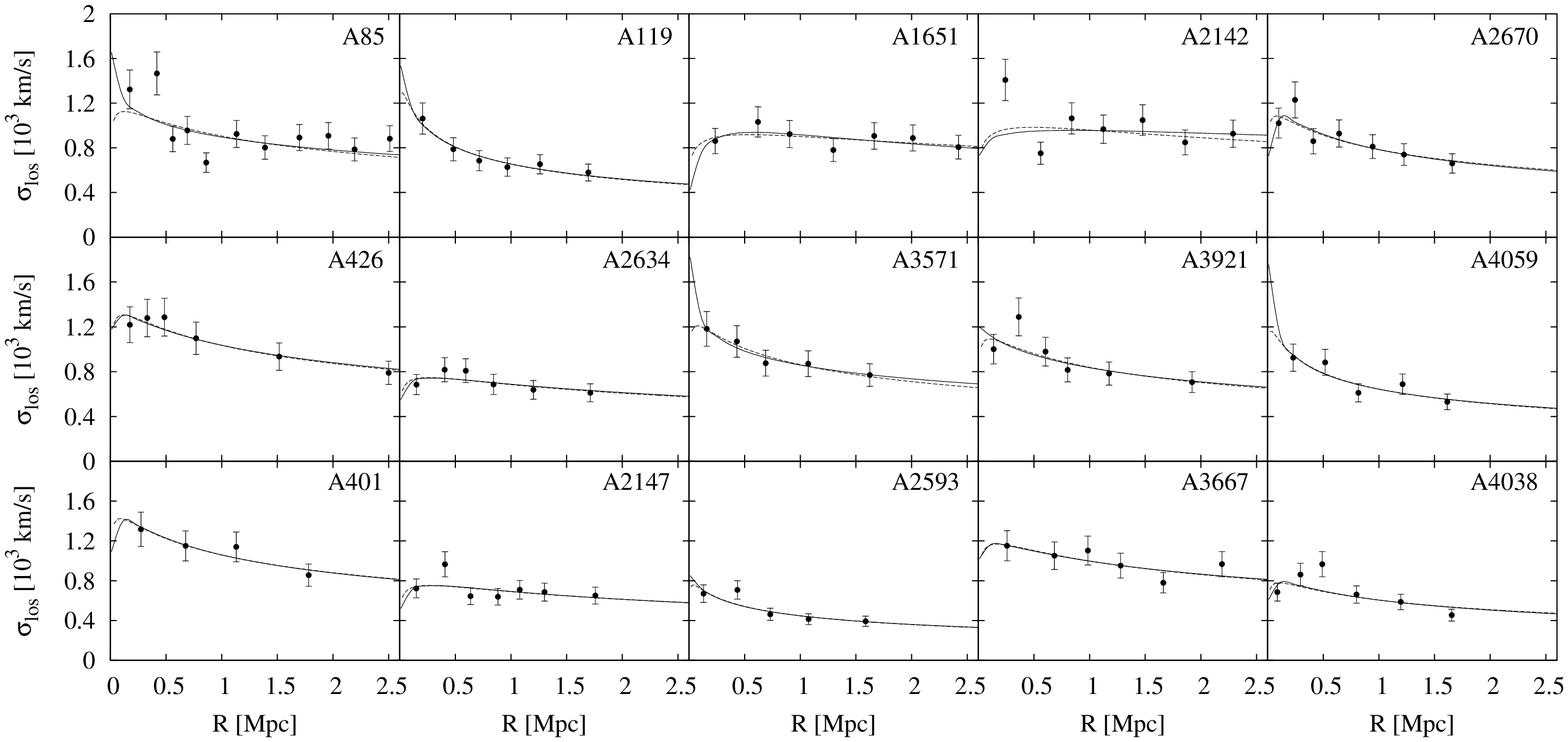}
\end{center}
\caption{Profiles of the velocity dispersion for 15 clusters listed in Table~\ref{clusters_int}.
In all cases the $v_{\rm max}$ method of interloper removal was applied except for A2147 and A4038
where the method based on $M_{P}/M_{VT}$ was used.
The dashed lines plot the best-fitting velocity dispersion profiles obtained for $\beta=0$, the
solid lines plot the solutions obtained when fitting both dispersion and kurtosis.
The best-fitting parameters are listed in Table~\ref{fitted}. }
\label{d_all}
\end{figure*}

\begin{figure*}
\begin{center}
    \leavevmode
    \epsfxsize=17cm
    \epsfbox[60 50 820 430]{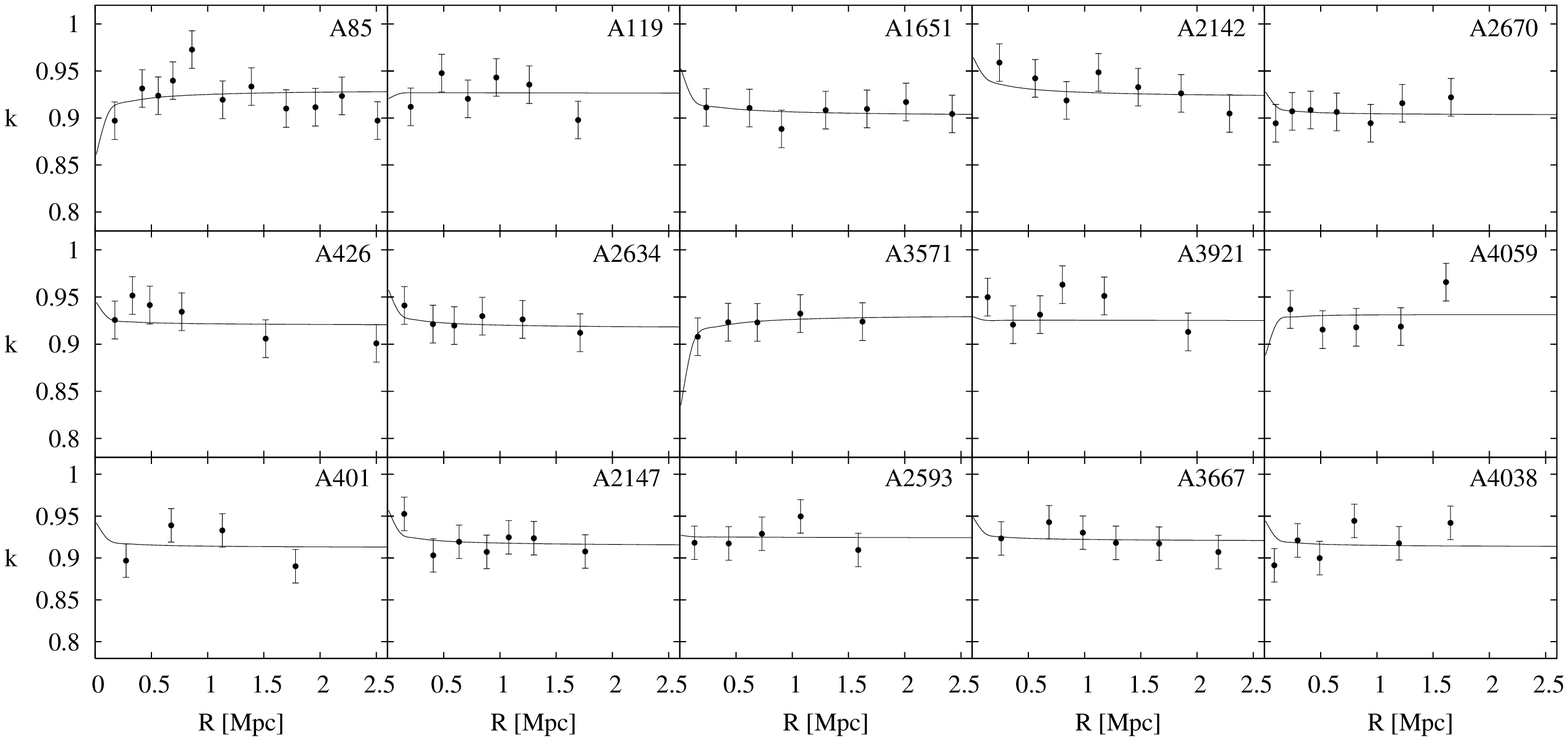}
\end{center}
\caption{Profiles of the
kurtosis variable $k = \left[ \log \left( 3 K/2.68 \right) \right]^{1/10}$
for the 15 clusters listed in Table~\ref{clusters_int}. The galaxy samples used are the same as for
Fig.~\ref{d_all}.
The solid lines plot the solutions of the Jeans equations obtained when
fitting both dispersion and kurtosis.}
\label{k_all}
\end{figure*}

\begin{table}
\begin{center}
\caption{Fitted parameters of 15 clusters from Table~\ref{clusters_int}
with 1$\sigma$ error bars. The parameters were obtained from fitting $\sigma_{\rm los}$ with $\beta=0$
and joint fitting
of $\sigma_{\rm los}$ and $\kappa_{\rm los}$ for arbitrary constant $\beta$. The galaxy samples for
A2147 and A4038 were obtained with the
$M_P/M_{VT}$ method, for the rest of the clusters with the $v_{\rm max}$ method.}
\begin{tabular}{lcccc}
cluster & $\beta$  & $M_v $               & $c$    & $\chi^2/N$ \\
        &          & $[10^{14} M_{\sun}]$ &        &            \\
\hline
A85     & 0                                        & $12.7^{+2.5}_{-2.1}$ & $11.3^{+26.7}_{-7.7}$ & 20.7/9 \\   \\
   & $0.66^{+0.17}_{-0.38}$  & $9.9^{+5.6}_{-2.2}$   & $2.3^{+12.7}_{-1.3}$   & 29.7/19  \\
\hline
A119    & 0                                        & $5.6^{+1.4}_{-1.2}$   & $47.0^{+>200}_{-37.0}$     & 1.1/4 \\   \\
   & $0.53^{+0.47}_{-1.05}$ & $6.5^{+4.5}_{-1.8}$  & $31.4^{+>200}_{-21.4}$ & 5.7/9     \\
\hline
A1651 & 0                                        & $13.1^{+3.1}_{-4.5}$   & $2.7^{+4.4}_{-1.7}$     & 2.2/5 \\   \\
 & $-1.95^{+1.95}_{-6.15}$ & $13.9^{+5.1}_{-3.7}$   & $7.1^{+8.2}_{-4.1}$ & 3.4/11     \\
\hline
A2142 & 0                                        & $15.8^{+3.7}_{-5.0}$   & $3.1^{+9.1}_{-2.1}$     & 12.9/5 \\   \\
 & $0.18^{+0.40}_{-0.68}$ & $11.2^{+7.3}_{-4.2}$   & $1.0^{+6.5}_{-0.0}$ & 17.0/11     \\
\hline
A2670 & 0                                        & $8.8^{+2.4}_{-1.9}$   & $16.2^{+21.8}_{-8.2}$     & 2.9/5 \\   \\
 & $-1.63^{+1.63}_{-5.47}$ & $7.6^{+2.3}_{-2.1}$   & $26.8^{+38.2}_{-13.8}$ & 4.8/11     \\
\hline
A426  & 0                                        & $17.5^{+4.7}_{-4.2}$   & $13.8^{+21.2}_{-7.8}$     & 0.8/4 \\   \\
 & $0.16^{+0.59}_{-1.31}$ & $18.1^{+6.2}_{-5.2}$   & $11.2^{+27.8}_{-7.6}$ & 5.7/9     \\
\hline
A2634 & 0                                        & $6.4^{+1.5}_{-1.4}$   & $4.3^{+6.1}_{-2.9}$     & 1.7/4  \\   \\
 & $-0.07^{+0.57}_{-1.60}$ & $6.5^{+2.4}_{-1.8}$   & $4.3^{+7.9}_{-2.8}$    & 2.4/9   \\
\hline
A3571 & 0                                        & $11.0^{+3.7}_{-2.9}$   & $18.5^{+54.5}_{-12.5}$     & 0.5/3 \\   \\
 & $0.71^{+0.29}_{-1.96}$ & $9.7^{+5.5}_{-4.7}$   & $2.9^{+71.1}_{-1.9}$ & 0.6/7     \\
\hline
A3921 & 0                                        & $10.5^{+2.9}_{-2.4}$ & $13.2^{+19.3}_{-8.0}$ & 3.8/4  \\   \\
 & $0.43^{+0.40}_{-0.76}$  & $10.9^{+3.3}_{-2.9}$  & $7.5^{+25.0}_{-5.0}$   & 11.4/9   \\
\hline
A4059 & 0                                        & $5.4^{+1.5}_{-1.2}$   & $34.9^{+>200}_{-26.9}$     & 2.6/3 \\   \\
 & $0.75^{+0.25}_{-1.05}$ & $6.9^{+3.4}_{-2.2}$   & $17.1^{+>200}_{-14.2}$ & 7.2/7     \\
\hline
A401 & 0                                        & $18.0^{+7.0}_{-5.2}$   & $18.3^{+108.7}_{-13.6}$     & 0.8/2 \\   \\
 & $-0.40^{+1.20}_{-4.10}$ & $17.0^{+7.7}_{-6.1}$   & $22.5^{+>200}_{-12.5}$ & 5.6/5     \\
\hline
A2147 & 0                                        & $6.5^{+1.4}_{-1.3}$   & $4.3^{+6.8}_{-3.0}$     & 4.7/5 \\   \\
 & $-0.27^{+0.70}_{-1.78}$ & $6.6^{+2.1}_{-1.7}$   & $5.4^{+9.3}_{-3.8}$    & 7.8/11   \\
\hline
A2593 & 0                                        & $2.4^{+0.6}_{-0.6}$   & $21.6^{+65.4}_{-14.1}$     & 3.0/3 \\   \\
 & $0.41^{+0.59}_{-1.36}$ & $2.6^{+1.1}_{-0.8}$   & $15.8^{+109.2}_{-9.8}$ & 5.6/7    \\
\hline
A3667 & 0                                        & $16.6^{+4.2}_{-3.6}$   & $9.1^{+21.9}_{-6.1}$     & 2.8/4 \\   \\
 & $0.16^{+0.65}_{-1.60}$ & $16.8^{+5.1}_{-4.3}$   & $7.5^{+56.5}_{-5.7}$ & 4.6/9     \\
\hline
A4038 & 0                                        & $4.8^{+1.2}_{-1.1}$   & $9.8^{+9.0}_{-5.0}$     & 8.8/4 \\   \\
 & $-0.33^{+1.03}_{-2.37}$ & $4.6^{+1.5}_{-1.2}$   & $12.7^{+14.3}_{-10.0}$ & 15.7/9     \\
\hline
\label{fitted}
\end{tabular}
\end{center}
\end{table}

The number of interlopers in a
sample for a given cluster depends on its particular
neighbourhood but also on the method of galaxy selection by
the observer (whether some photometric criteria were used).
The fraction of interlopers can therefore be very different
and is not so important. More important is their velocity
distribution. If there are only a few interlopers but with
highly discrepant velocities they can also strongly bias
the inferred parameters of mass distribution. However, such interlopers are typically
easily removed by simple methods like the rejection of the $3\sigma$ outliers. In A576 the
interlopers are more uniformly distributed in velocity which requires the use of more
sophisticated methods like the ones presented here (still, the methods are simpler in
application than the caustic method of Rines et al. 2000). Since regular
clusters are scarce and the problem of
contamination by interlopers is common among them we think even those most contaminated
should be studied. We demonstrated that in many such cases proper removal of interlopers can
lead to reliable determination of galaxy membership.

The general picture that emerges from our study of 16 clusters is that even in the case of
significant contamination different interloper removal schemes can yield similar results increasing
our confidence in the final galaxy samples. In the case when two or more schemes (usually
the methods based on $v_{\rm max}$ and $M_{P}/M_{VT}$) give similar results we recommend the use
of $v_{\rm max}$ since it converges and is independent of any assumptions. There are cases however
with strong substructure visible in the velocity diagrams when only the $M_{P}/M_{VT}$ method
seems to give results which clearly separate the subclusters from the main body of the cluster of
interest. We recommend this method for such cases.

Following these conclusions we have selected final galaxy samples for our 15 clusters presented
in Fig.~4a-c of the previous section:
we used the $M_{P}/M_{VT}$ method for A2147 and A4038 and the $v_{\rm max}$ method for all the remaining
clusters. The resulting profiles of the velocity moments are shown in Fig.~\ref{d_all} and \ref{k_all}.
As for A576, we fitted the data for the velocity dispersion and kurtosis to the solutions of the
Jeans equations with $\beta=$ const estimating the virial mass, concentration
and anisotropy parameter. The best-fitting parameters together with 1$\sigma$ error bars are
given in Table~\ref{fitted}. We have also listed there for comparison the parameters obtained
previously for the same galaxy samples from fitting velocity dispersion alone assuming $\beta=0$.
The fitted profiles are shown in Fig.~\ref{d_all} and \ref{k_all} with solid lines
(velocity dispersion profiles obtained for $\beta=0$ are plotted in Fig.~\ref{d_all}
with dashed lines which in most cases overlap almost exactly with the solid lines).

\begin{figure}
\begin{center}
    \leavevmode
    \epsfxsize=7.2cm
    \epsfbox[90 10 290 210]{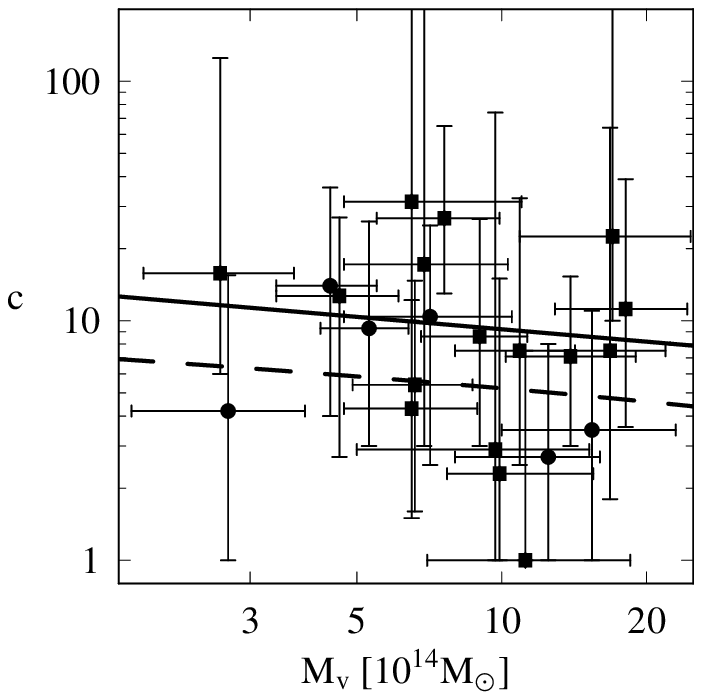}
\end{center}
\caption{The mass-concentration relation. The points show the best-fitting
$M_v$-$c$ pairs for 22 clusters with $1\sigma$ error bars. The circles correspond
to the 6 clusters studied in {\L}okas et al. (2006a), the squares to the 16 new clusters
studied here. The dashed line
plots the prediction from the $N$-body simulations by Bullock et al. (2001) while
the solid line shows the best fit to the data points.}
\label{cm}
\end{figure}

These new results allow us to extend the mass-concentration relation given in {\L}okas et al. (2006a)
to the total of 22 clusters. The relation is shown in Fig.~\ref{cm} with circles plotting the
parameters for the 6 clusters studied in {\L}okas et al. and squares for the 16 new clusters studied
here. The dashed line shows the approximation of the $c(M_v)$ relation calculated from the
toy model proposed by Bullock et al. (2001) which reproduces well the properties of
a large sample of haloes found in their $N$-body simulations. The fit of a linear
relation of the form $\log c = a \log [M_v/(10^{14} M_{\sun})] + b $ to the data gives
best-fitting parameters significantly different from those in {\L}okas et al.:
$a=-0.17 \pm 0.65$ and $b=1.14 \pm 0.62$ (at 68 percent
confidence level). This best-fitting relation is shown in Fig.~\ref{cm} as a solid line: the slope
of the fitted relation agrees well with the one from  $N$-body simulations, but our
concentrations are shifted towards higher values.
The results agree within errors with other determinations of the mass-concentration relation
(e.g. Pointecouteau, Arnaud \& Pratt 2005; Vikhlinin et al. 2006; Mandelbaum et al. 2006;
Rines \& Diaferio 2006; Schmidt \& Allen 2006; Buote et al. 2006). Interestingly, our higher typical
concentration values agree with other recent determinations for individual clusters (e.g.
Broadhurst et al. 2005).

Note that the profiles of the velocity moments in Fig.~\ref{d_all} and \ref{k_all}
determined from the cleaned galaxy samples are very regular,
with expected behaviour: the dispersion profile decreases slowly at larger distances while the kurtosis
profile is rather flat. The profiles are well behaved even in the case of A119, A2634 and A2147 in
spite of the fact the surface brightness distributions of the X-ray gas are least regular in these
clusters. From this point of view there is no difference between these originally contaminated
clusters and
those presumably more regular ones studied in {\L}okas et al. (2006a). The difference between these
two groups manifests itself however in the fitted parameters. While in the previous study 4 out of
6 clusters had the best-fitting anisotropy parameter $|\beta|<0.2$, for the present sample this is
only the case for 5 clusters out of 16. Two of the clusters (A1651 and A2670) have
$-2 < \beta < -1.6$ corresponding to
$\sigma_{\theta}/\sigma_r=1.6-1.7$, indicating significant tangential anisotropy, while three clusters
(A85, A3571 and A4059) have $\beta \approx 0.7$ corresponding to $\sigma_r/\sigma_{\theta}=1.8$,
indicating significant radial anisotropy. This may point towards a more perturbed state of the present
sample. Still, all clusters have $\beta$ consistent with isotropy at 1$\sigma$ level, except for A85.

A85 seems to be the most suspicious case in the present sample. While the removal of interlopers
would suggest no problems in the analysis (all three methods yield very similar results in terms
of the number of rejected interlopers as well as the fitted mass and concentration, see
Table~\ref{clusters_int} and Fig.~\ref{clusters1}), while fitting
both velocity moments we arrive at rather different parameters, lower mass and concentration and more
radial orbits. The profiles of the velocity moments are also not typical: the first two dispersion
data points are much higher than the rest and the kurtosis points lower. This may point at departures
from equilibrium which are indeed confirmed by Kempner, Sarazin \& Ricker (2002) and
Durret, Lima Neto \& Forman (2005) who studied the
X-ray data from Chandra and XMM-Newton and discovered a subcluster about to merge with A85 as well as
evidence for past merging events. Note, however, that the mass estimated from X-ray data by Durret et al.
within $r<1.3$ Mpc, $M_X = 4.3 \times 10^{14} M_{\sun}$, is in very good agreement with our mass within
this radius $M = 4.5 \times 10^{14} M_{\sun}$.

Reiprich \& B\"ohringer (2002) used the ROSAT X-ray data to estimate the masses of all clusters studied
here except for A2593 and A2670. In all cases our estimates agree with their findings within errors,
only for A426, A3667 and A4038 our masses are about a factor of two higher.
Our mass estimates also agree within errors with the ones found from the virial theorem
by Girardi et al. (1998) in the case of A85, A119, A401, A426, A576, A2593, A2634, A2670, A3571
and A3667. For A2142 our mass estimate is more than twice lower than the one of Girardi et al. The
result in this case may be strongly dependent on the sample selection because the velocity diagram
of the cluster is rather irregular.

For A3921 our estimate of the virial mass
is almost four times higher than the one of Girardi et al. which is probably due to their rather small
sample of only 29 member galaxies while our sample contains 197 members. Referring to X-ray studies
with XMM-Newton
(Belsole et al. 2005) we find that there are signatures of recent merger in the form of a bar-like
structure of enhanced temperature. Belsole et al. (2005) estimate the mass to be $M_X = 2 \times
10^{14} M_{\sun}$ within $r<0.8$ Mpc while within this distance we get $M = 4.2 \times 10^{14}
M_{\sun}$, this time a value a factor of two higher. Note however, that our estimate is in very good
agreement with the older ROSAT data (Reiprich \& B\"ohringer 2002) which give $M_v = 11 \times 10^{14}
M_{\sun}$. Going back to
the velocity diagram for A3921 in Fig.~\ref{clusters2} we see that there are some doubtful
galaxies (e.g. the chain at 2000 km s$^{-1}$) which are not rejected by our algorithms but which
may not be cluster members and may in fact overestimate the true cluster mass.

A2147 is a member of the Hercules supercluster which obviously makes the assignment of galaxies
to each cluster more difficult (Tarenghi et al. 1980; Barmby \& Huchra 1998). Together with A4038 it
is the only cluster for which neighbours are well visible and we believe the $M_{P}/M_{VT}$ method
of interloper removal is more reliable than $v_{\rm max}$. In the case of A2147 our mass estimate is
about twice lower than the one based on velocity dispersion by Barmby \& Huchra (1998). The difference
is obviously due to the different selection of galaxies, for example they did not
remove the group at $-1300$
km s$^{-1}$ (see Fig.~\ref{clusters3}). In the case of A4038 we agree with the rough estimate of the
cluster mass based on galaxies in the core by Green, Godwin \& Peach (1990).

We have demonstrated that after careful removal of interlopers a significant number of galaxy clusters
can be made eligible for Jeans analysis. Using the example of A576 we have shown in detail how
the different most reliable methods of cleaning the kinematic samples of interlopers
work on real data. We illustrated the complexity of the problem
with examples of other clusters displaying different degrees of contamination and different
distributions of interlopers. The results of the final dynamical analysis of these clusters are
not significantly different from those obtained for more regular clusters increasing our confidence
in the applicability of Jeans formalism.

\section*{Acknowledgements}

We are grateful to
S. Gottl\"ober, A. Klypin, G. A. Mamon, M. Moles and F. Prada for discussions.
This research has made use of the NASA/IPAC Extragalactic Database (NED) which is
operated by the Jet Propulsion Laboratory, California Institute of Technology,
under contract with NASA. RW acknowledges the summer student program at Copernicus Center.
This work was partially supported by the Polish Ministry of Science and Higher Education
under grant 1P03D02726.

\end{document}